\documentclass[pra,twocolumn,amsmath,amssymb,letterpaper,superscriptaddress,nofootinbib,longbibliography]{revtex4-2}
\usepackage{graphicx}
\usepackage{amsmath}
\usepackage{amsfonts}
\usepackage{amssymb}
\usepackage{ulem}
\usepackage[colorlinks,linkcolor=blue,filecolor=blue,urlcolor=blue,citecolor=blue]{hyperref}

\begin{document}
	
	\title{Many-body non-Hermitian skin effect under dynamic gauge coupling}
	
	\author{Haowei Li}
	\affiliation{CAS Key Laboratory of Quantum Information, University of Science and Technology of China, Hefei 230026, China}
	\author{Haojie Wu}
	\affiliation{Hefei National Laboratory for Physical Sciences at the Microscale and Department of Modern Physics,
		University of Science and Technology of China, Hefei 230026, China}
	\author{Wei Zheng}
	\email{zw8796@ustc.edu.cn}
	\affiliation{Hefei National Laboratory for Physical Sciences at the Microscale and Department of Modern Physics,
		University of Science and Technology of China, Hefei 230026, China}
	\affiliation{CAS Center For Excellence in Quantum Information and Quantum Physics, Hefei 230026, China}
	\affiliation{Hefei National Laboratory, University of Science and Technology of China, Hefei 230088, China}
	\author{Wei Yi}
	\email{wyiz@ustc.edu.cn}
	\affiliation{CAS Key Laboratory of Quantum Information, University of Science and Technology of China, Hefei 230026, China}
	\affiliation{CAS Center For Excellence in Quantum Information and Quantum Physics, Hefei 230026, China}
	\affiliation{Hefei National Laboratory, University of Science and Technology of China, Hefei 230088, China}
	
	\begin{abstract}
		We study an atom-cavity hybrid system where fermionic atoms in a one-dimensional lattice are subject to a cavity-induced dynamic gauge potential.
		The gauge coupling leads to highly-degenerate steady states in which the fermions accumulate to one edge of the lattice under an open boundary condition.
		Such a phenomenon originates from the many-body Liouvillian superoperator of the system, which, being intrinsically non-Hermitian, is unstable against boundary perturbations and manifests the non-Hermitian skin effect.
		Contrary to the single-body case, the steady state of a multi-atom system is approached much slower under the open boundary condition,
		as the long-time damping of the cavity mode exhibits distinct rates at different times.
		This stage-wise slowdown is attributed to the competition between light-assisted hopping and the dynamic gauge coupling, which significantly reduces the steady-state degeneracy under the open boundary condition, as distinct hosts of quasi-steady states dominate the dynamics at different time scales.
	\end{abstract}
	
	\maketitle
	
	\section{Introduction}
	
	\begin{figure}[tbp]
		\includegraphics[width=0.45\textwidth]{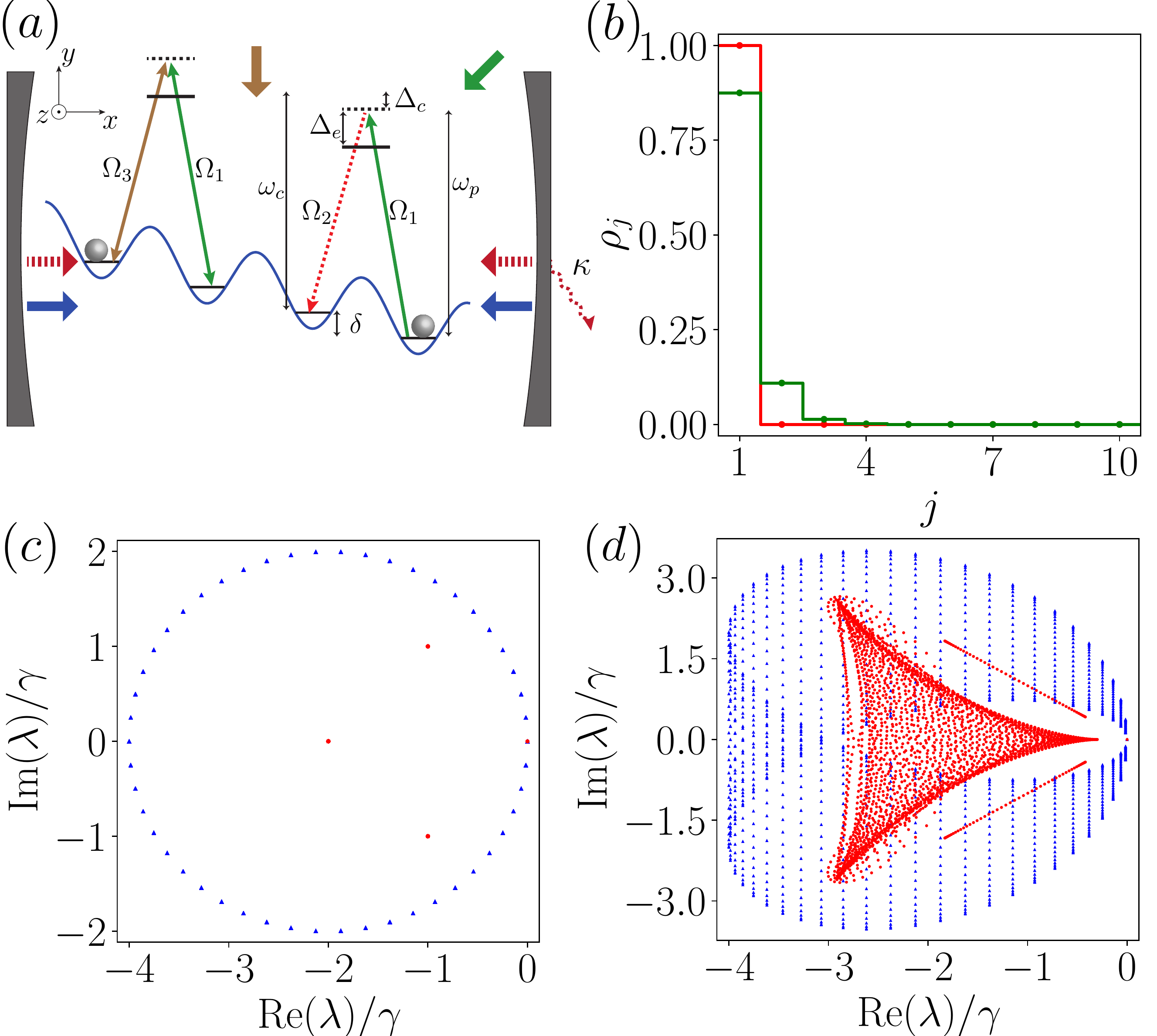}
		\caption{Schematic of the setup and single-particle NHSE.
			(a) Spinless fermions are subject to Raman-assisted hoppings along a one-dimensional lattice. One of the Raman processes consists of the cavity mode (red dashed arrow, with a frequency $\omega_c$),
			and the cavity pump laser (green solid arrow, with a frequency $\omega_p=\omega_c+\delta$), with a two-photon detuning of $\Delta_e$.
			The other Raman process is generated by the pump and an additional Raman laser (brown solid arrows).
			The lattice is tilted with an on-site detuning $\delta$ to switch off direct hopping.  A large cavity loss rate $\kappa$ is assumed.
			(b) Spatial distribution of the single-particle steady states for $s=0$ (red) and $s/\gamma=0.5$ (green) under OBC.
			(c) Single-particle Liouvillian spectrum on the complex plane for $s=0$ under PBC (blue) and OBC (red), respectively.
			(d) Single-particle Liouvillian spectrum for $s/\gamma=0.5$, under PBC (blue) and OBC (red), respectively.
			For calculations in (b)(c)(d), we take $L=50$ and $N=1$.
		} \label{Fig1}
	\end{figure}
	
	Gauge fields are a central topic in modern physics---the elegant formulation of quantum matter interacting with gauge fields underlies
	many distinct physical settings, ranging from high-energy physics~\cite{hep1,hep2,hep3,hep4} to strongly-correlated quantum materials~\cite{frachall,fisher00,spinliquid} .
	The recent implementation of synthetic gauge potentials in ultracold atomic gases~\cite{syngf1,syngf2,syngf3,syngf4,xiongjun16, chin18,DDGF@Esslinger2019,Tarruell2022,z2gauge,Bloch2019,yuan20,Jendrzejewski2020} sheds new light on the subject---
	not only does a rich variety of gauge potentials become experimentally accessible, its interplay with strong interaction and dissipation can be investigated in a controlled manner.
	Within this context, atom-cavity hybrid systems offer a particularly intriguing avenue~\cite{Esslinger2010,WuHaiBing2021,esslinger12,cqed17,cqed18,esslinger22}. The cavity mode, by inducing a dynamic gauge potential, gives rise to long-range interatomic interactions~\cite{esslinger16}, which, combined with the back action of cavity dissipation, can lead to exotic far-from-equilibrium dynamics and many-body steady states~\cite{thorwart15, zheng16,morigi13,strack14,jiansong14, kollath20,Ueda2018,Esslinger2019,DTC2021,CTC2022,ZW2023}.
	
	For cold atoms coupled to a lossy cavity, the dynamics is driven by the Liouvillian superoperator~\cite{molmer92,carmichael93,uedaadvphys20,weimer21}.
	Its intrinsic non-Hermiticity suggests that unique features of non-Hermitian Hamiltonians, such as the parity-time symmetry~\cite{bender98,elganainy18}, criticality and topology associated with the exceptional points~\cite{alu19,nori19,uedacritical}, and the non-Hermitian skin effects (NHSE)~\cite{wang1d,wang2d,murakami19,thomale19,yi19,fang20,longhi19,longhi20,sato20,longhiprr19,wang19,yi21,ameba,skinrev1,skinrev2}, can impact the open-system dynamics.
	Of particular interest here is the NHSE, which originates from the instability of non-Hermitian matrices to boundary perturbations, and manifests in the accumulation of eigenstates toward the boundary under the open boundary condition (OBC).
	The NHSE is found to have dramatic influence on the system's band and spectral topology~\cite{fang20,sato20,ameba}, as well as the spectral symmetry~\cite{longhioptl19,xue21,chen21} and dynamics~\cite{wang19,yi21,longhiprr19,xueskin,yanskin,xuedynamics}.
	
Some of these influences carry over to non-Hermitian models within the many-body context, as the interplay of the NHSE and many-body interaction can have impact on the system's topology~\cite{inttopo1,inttopo2,intcomp1,intcomp2} and localization~\cite{intmbl1,intmbl2}.
However, the manifestations of NHSE in many-body quantum open systems, particularly in the experimentally relevant hybrid atom-cavity configurations, remain unexplored.
	
	In this work, we study the emergence and consequence of NHSE in fermionic atoms under a cavity-induced dynamic gauge potential.
	The Liouvillian spectra of this many-body open system, while sensitive to boundary conditions, exhibit rich structures that are distinct from the single-particle case, leading to boundary-dependent damping dynamics.
	Under the OBC, for instance, the steady-state population accumulates toward the boundary. The steady-state degeneracy can be significantly reduced compared to that under the periodic boundary condition (PBC), as hosts of quasi-steady eigenmodes emerge in the Liouvillian spectrum.
	The Liouvillian eigenvalues of these quasi-steady modes have power-law scalings with respect to the system parameters,
	giving rise to a significant slowdown of the steady-state-approaching dynamics.
	Intriguingly, the power-law scaling of the Liouvillian egienvalues are determined by the localization properties of the corresponding eigenmodes, such that the system becomes more and more localized as different groups of quasi-steady modes take turns to dominate the long-time dynamics.
	Our results reveal the non-trivial effects of NHSE in a many-body quantum open system, and highlight the atom-cavity hybrid system as an experimentally accessible setup where non-Hermitian physics can be investigated.

The work is organized as follows. In Sec.~II, we present the setup of the atom-cavity hybrid system, and reveal the NHSE on the single-body level.
We proceed in Sec.~III by considering the cases with more than one atoms, and discuss the distinct features of the NHSE beyond the single-body level.
The impacts of the many-body NHSE on the steady-state degeneracy and long-time dynamics are then discussed in detail in Secs.~IV and V, respectively. We summarize in Sec.~VI.
	
	\begin{figure*}[tbp]
		\includegraphics[width=1\textwidth]{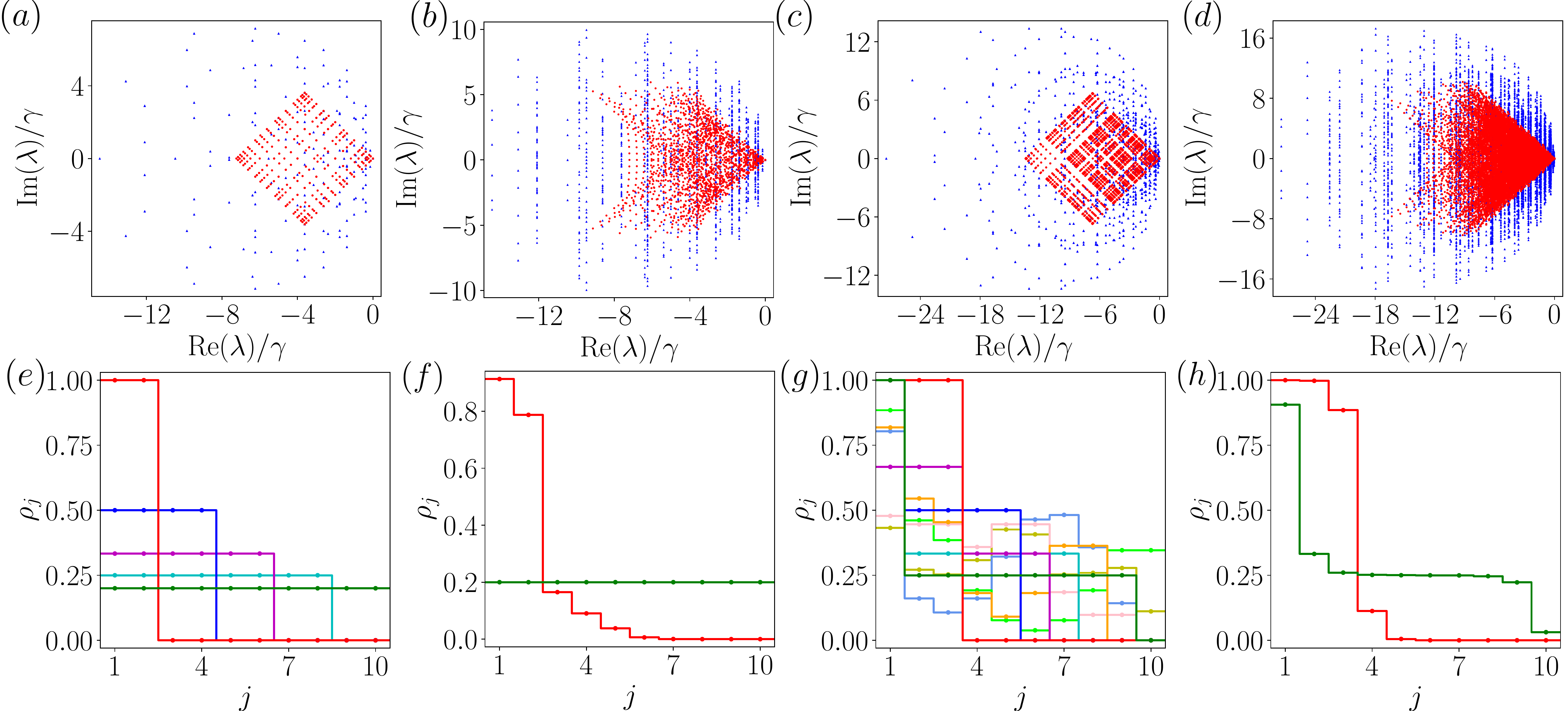}
		\caption{NHSE with more than one fermions. (a)(b)(c)(d) Liouvillian spectra for a fixed lattice length $L=10$
			with (a)(b) $N=2$ and (c)(d) $N=3$ fermions, respectively. The blue (red) spectra are calculated under the PBC (OBC).
			(e)(f)(g)(h) Spatial density distribution of typical steady states under OBC, for (e)(f) $N=2$ and (g)(h) $N=3$ fermions, respectively.
			We take $s=0$ in (a)(c)(e)(g), and $s/\gamma=0.5$ in (b)(d)(f)(h).
		} \label{Fig2}
	\end{figure*}
	
	\section{Model and single-body NHSE}
	We consider spinless fermions in a quasi-one-dimensional lattice with cavity-assisted hopping [see Fig.~\ref{Fig1}(a)].
	The lattice potential is along the $x$ direction, and tightly confined in the other directions.
	It is further tilted by a magnetic gradient, so that the detuning $\delta$ between neighboring sites suppresses direct inter-site hopping.
	Two Raman-assisted-hopping processes are introduced. One is mediated by the cavity pump [green solid in Fig.~\ref{Fig1}(a)] and the cavity mode (red dashed), where the two-photon detuning $\Delta_e\ll \delta$.
	The other is generated by the pump beam and an additional Raman laser (brown solid arrow).
	Given a lossy cavity, dynamics of the fermions adiabatically follows that of the cavity field,
	and the fermions are effectively subject to a cavity-dependent dynamic gauge potential.
	In the following, to differentiate the two processes, we address them as the dynamic gauge coupling and the Raman-assisted hopping, respectively.
	
	Eliminating the cavity mode and taking the tight-binding approximation~\cite{zheng16,ritsch05,jiansong14}, the atomic density matrix $\rho$ is governed by the Lindblad master equation (we take $\hbar=1$)
	\begin{align}
		\frac{d {\rho}}{d t}=\mathcal{L}[\rho]=-\mathrm{i}[\hat{H}, {\rho}]+ \gamma\left(2 \hat{K} {\rho} \hat{K}^{\dagger}-\left\{\hat{K}^{\dagger} \hat{K}, {\rho}\right\}\right),
	\end{align}
	where $\mathcal{L}$ is the Liouvillian superoperator, and
	\begin{align}
		\hat{H}=-\frac{\Delta_c}{\kappa}\gamma \hat{K}^\dagger \hat{K}+s(\hat{K}+\hat{K}^\dagger),\label{eq:effH}
	\end{align}
	with $\gamma=\kappa \lambda^2/(\Delta_c^2+\kappa^2)$. Throughout the work, we take $\Delta_c/\kappa=1$ for numerical calculations.
	The jump operator $\hat{K}=\sum_j^{L-1} \hat{c}^{\dagger}_{j+1}\hat{c}_j$, $s$ is the amplitude of the Raman-assisted hopping, where $\hat{c}_j$ ($\hat{c}^\dag_j$) annihilates (creates) a fermion on site $j$, and $L$ is the total number of sites.
	Notably, the dynamic gauge coupling gives rise to cavity-mediated long-range interactions, captured by the first term in Eq.~(\ref{eq:effH}).
	
	Insights on the long-time dynamics can be gained from the Liouvillian spectrum $\lambda$, which satisfies the eigen equation $\mathcal{L}[\rho]=\lambda\rho$, with $\text{Re}\lambda\leq 0$ for the purely dissipative system considered here. The steady states correspond to eigenstates with $\lambda=0$, while $\text{Re}\lambda$ of other eigenmodes contribute to the system's relaxation toward the steady states.
	
	An outstanding feature of the hybrid system is the sensitive dependence of the steady states on the boundary condition.
	In the simplest case where a single atom is coupled to the cavity under PBC and with $s=0$, the steady state exhibits a uniform density distribution but a directional bulk current. Under the OBC, however, the bulk current vanishes and the steady state corresponds to a fully localized atom on the leftmost site [red in Fig.~\ref{Fig1}(b)].
	When the Raman-assisted hopping is switched on with $s\neq 0$, it competes with the localizing effect of the dynamic gauge coupling, leading to a partially localized steady state [green in Fig.~\ref{Fig1}(b)].
	While configurations with $s=0$ were investigated previously in the context of non-equilibrium dynamics~\cite{zheng16},
	these properties are reminiscent of the chiral current (under PBC)~\cite{longhiprr19,wang19,xueskin,yanskin}  and boundary localization (under OBC)~\cite{wang1d} of the recently discovered NHSE.
	
	The presence of NHSE is more transparent from the Liouvillian spectra [see Fig.~\ref{Fig1}(c)(d)].
	Under PBC, the spectra form looped structures on the complex plane [blue in Fig.~\ref{Fig1}(c)(d)], consistent with the well-known spectral topology of the NHSE~\cite{fang20,sato20}.
	By contrast, the spectra collapse under OBC, yielding a larger Liouvillian gap. Here the Liouvillian gap is defined as the minimum $|\text{Re}\lambda|$ other than $0$.
	Consequently, the steady states are approached much faster under OBC, than in the case of PBC.
	The NHSE here derives from the sensitivity of the Liouvillian eigenmodes and eigenspectrum to boundary conditions, which is reflected in the relations $[\hat{K},\hat{K}^\dag]=0$ under PBC, and $[\hat{K},\hat{K}^\dag]\neq0$ under OBC.
	Since these relations also hold in the many-body case, one expects the NHSE should persist when multiple fermions are coupled to the cavity.
	
	\begin{figure}[tbp]
		\includegraphics[width=0.5\textwidth]{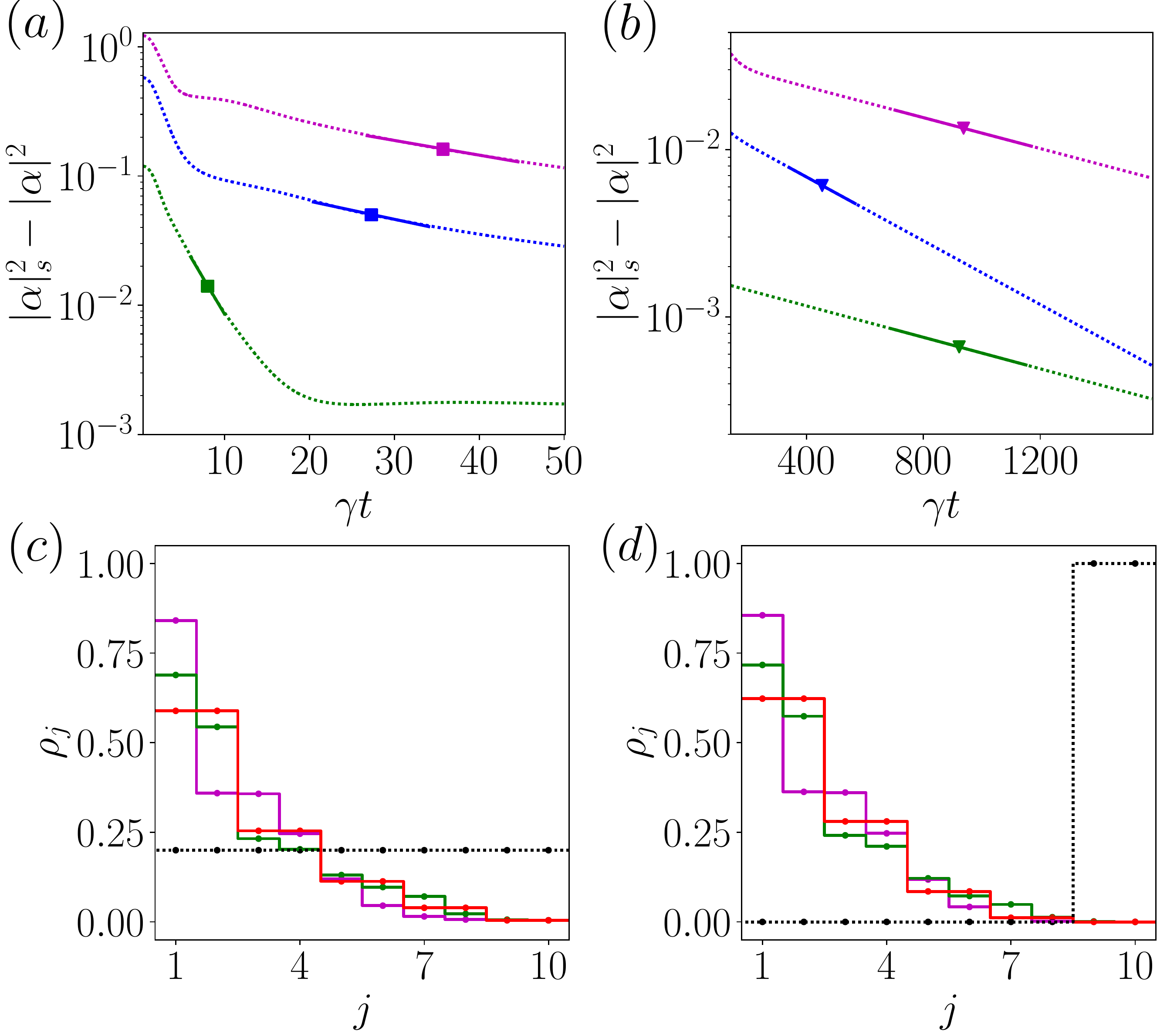}
		\caption{(a)(b) Long-time damping of the cavity field $|\alpha|^2$ under OBC, for $s=0.5,1.1, 1.6$ (green, blue, purple), respectively.
			We show $|\alpha^2|$ relative to the steady-state value $|\alpha_s|^2$, which is estimated using $|\alpha|^2$ at $\gamma t=10^5$.
			(c)(d) Atomic density distribution for $s/\gamma=0,0.5,1.6$ (red, green, purple), respectively, at time $\gamma t=10^4$.
			The initial states are given by a diagonal density matrix with even on-site distribution in (a)(b)(c), and a localized Fock state in (d), their spatial distributions indicated by the black dashed lines in (c)(d).
			For all figures, $L=10$ with $N=2$ fermions.
			The markers in (a)(b) indicate the typical time scales $t_q$ at which a given group of quasi-steady modes dominates the dynamics (see Fig.~\ref{Fig4}).
		} \label{Fig3}
	\end{figure}
	
	\begin{figure}[tbp]
		\includegraphics[width=0.5\textwidth]{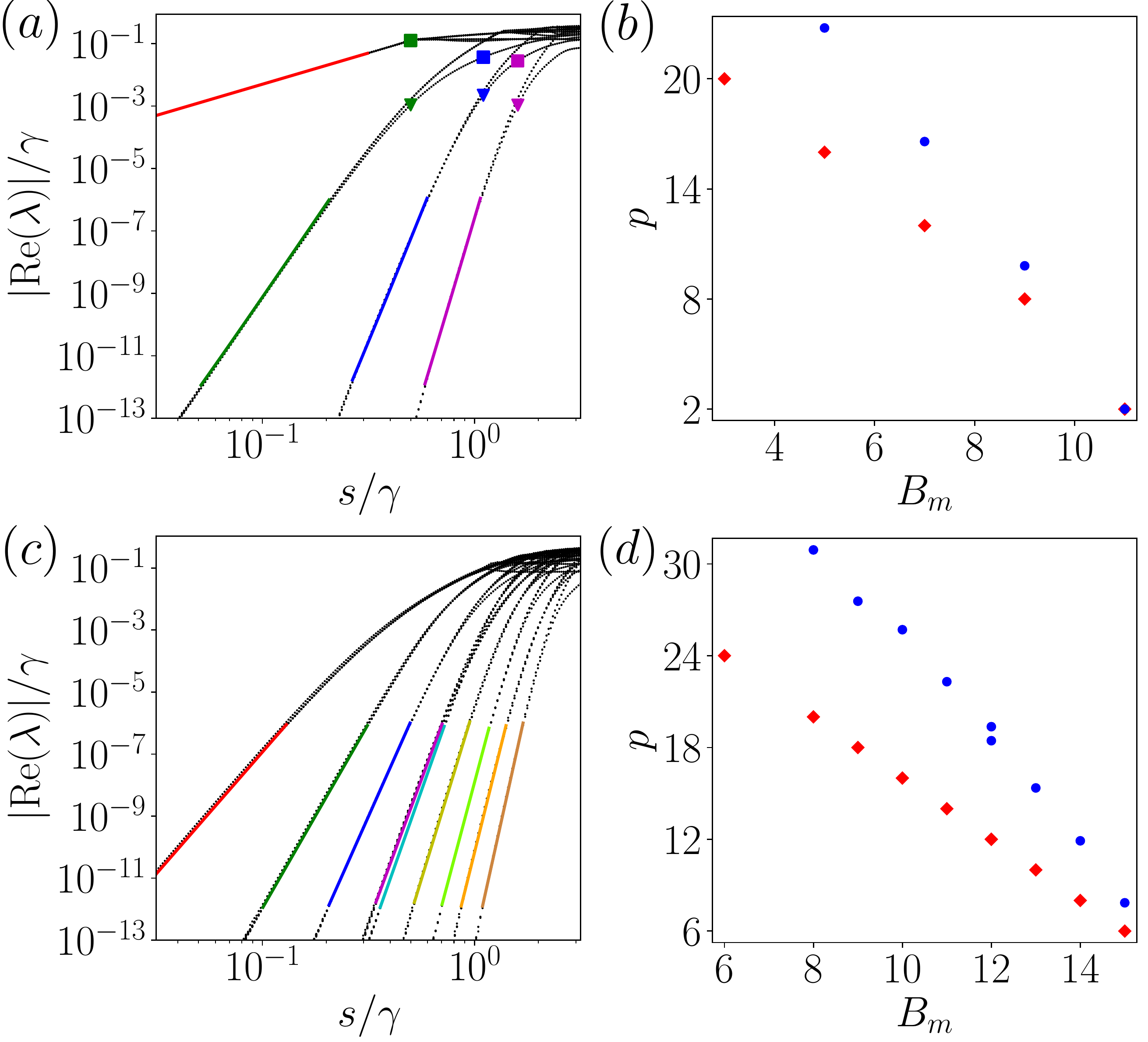}
		\caption{
			(a)(c) Real components of the Liouvillian spectra for the quasi-steady modes under OBC, as $s/\gamma$ increases in the vicinity of $s=0$. Only eigenmodes with eigenvalues close to $\text{Re}\lambda=0$ are shown. The green, blue and purple markers are taken at $s/\gamma=0.5,1.1,1.6$, respectively. Their shapes correspond to different branches of quasi-steady modes, which dominate at different time scales (see Fig.~\ref{Fig3}).
			The marked time scales shown in Fig.~\ref{Fig3} are estimated using $t_q=1/\text{Re}\lambda$ for the corresponding markers.
			(b)(d) Power-law exponent $p$ as a function of $B_m$. Calculations from numerical fit [over solid-line segments in (a)(c)] are indicated by blue dots, and the analytically obtained lower bounds are shown in red diamond. We take $L=10$, and (a)(b) $N=2$ and (c)(d) $N=3$ fermions, respectively.
		} \label{Fig4}
	\end{figure}
	
	\section{Many-body NHSE}
	
	As shown in Fig.~\ref{Fig2}(a)(b)(c)(d), in the multiple-fermion case, the Liouvillian spectra differ dramatically under different boundary conditions, while the steady states of the system become localized toward a boundary under OBC [see Fig.~\ref{Fig2}(e)(f)(g)(h)].
	However, compared to the single-body case, there are important differences.
	First, the loop structure of the spectrum under PBC disappears under finite $s$, suggesting the lack of spectral topology in the many-body case.
	Second, the Liouvillian gap under OBC is smaller than that under PBC, reversing the situation in the single-fermion case, and affecting the long-time dynamics, as we show below.
	Last but not least, while the steady-state degeneracy is typically quite large under OBC at $s=0$, it is reduced significantly under a finite $s$.
	The lifting of the steady-state degeneracy is accompanied by the emergence of eigenmodes that lie close to $\lambda=0$ and dominate the long-time dynamics. In the following, we address these eigenmodes as quasi-steady modes.
	
	\section{Steady-state degeneracy}
	
	To understand the steady-state degeneracy, we adopt a Fock basis $\{|j_1,j_2,\cdots, j_N\rangle\}$, where $j_i\in [1,L]$ indicates the site index occupied by the $i$th fermion, with the understanding that each basis state is properly anti-symmetrized.
	Here the lattice sites are labeled in ascending order from left to right.
	The Hilbert space is then divided into different subspaces $\mathcal{H}_B$, each with a fixed sum of occupied-site index $B=\sum^N_{i=1} j_i$.
	Such an index quantitatively characterizes the extent of localization---states localized to the left (right) boundary possess small (large) index.
	Since the jump operator $\hat{K}$ transforms a state in the subspace $\mathcal{H}_B$ to one in $\mathcal{H}_{B-1}$, its matrix in the Fock basis is block-off-diagonal.
	
	In the case of PBC, since $[\hat{K},\hat{K}^\dag]=0$, $\hat{K}$ and $\hat{K}^\dag$ have common eigenstates, and complex-conjugate eigenvalues. It is then straightforward to show that the steady states of the Liouvillian $\mathcal{L}$ can be constructed from the eigenstates of $\hat{K}$ (regardless of the value of $s$). Specifically, denoting the eigenstates of $\hat{K}$ as $\{|\phi_{n\beta}\rangle\}$, where
	$n$ and $\beta$ respectively label the eigenvalue and the corresponding degenerate eigenstate,
	the steady-state density matrix can be constructed as arbitrary superpositions of $\{|\phi_{n\beta}\rangle\langle |\phi_{n\beta'}|\}$. The steady-state degeneracy is therefore $\sum_n \beta_n^2$, where $\beta_n$ is the degeneracy of the $n$th eigenvalue.
	Since $\sum_n \beta_n^2\geq \frac{L!}{N!(L-N)!}$, the latter being the dimension of the full Hilbert space, the steady states are highly degenerate under PBC, with the degeneracy dependent only on $L$ and $N$.

	Under OBC by contrast, $[\hat{K},\hat{K}^\dag]\neq 0$, and the steady-state degeneracy becomes $s$-dependent.
	When $s=0$, the steady states can only be constructed from the dark states of $\hat{K}$.
	From the block-off-diagonal nature of the $\hat{K}$ matrices, one can deduce that the number of dark states is at least $G_m$, where $G_m$ is defined as the maximum dimension of all possible $\mathcal{H}_{B}$. For all cases considered in this work, we have numerically checked that the lower bound $G_m$ gives the exact number of dark states.
	Further, in the limit $L,N\gg 1$, we apply the central limit theorem to find that
	$G_m\approx \sqrt{\frac{6}{\pi N(L-N)(L+1)}}\frac{L!}{N!(L-N)!}$ (see Appendix B).
	The corresponding steady-state degeneracy is then $G_m^2$, which already provides an accurate estimation for $N=2,3$ considered in this work.
	
	Under finite $s$, we find that, for any $\rho_0$ with $[\hat{K},\rho_0]=[\hat{K}^\dag,\rho_0]=0$,
	the steady states $\rho_s$ can be generated through (see Appendix C)
	\begin{align}
		\rho_s=\sum_n\sum^{n}_{q=0} A_+^{-q}A_-^{-(n-q)}\hat{K}^q(\hat{K}^\dagger)^{n-q}\rho_0,
	\end{align}
	where $A_\pm=\frac{\mp is}{(1\pm i\Delta_c/\kappa)\gamma}$. Since $\hat{K}^q=(\hat{K}^\dag)^q=0$ for $q>NL$, the summation over $n$ has finite terms.
	Exhausting all possible forms of $\rho_0$, one can prove that the steady-state degeneracy is significantly reduced to
	$1+\frac{1+(-1)^L}{2}\lfloor \frac{N}{2}\rfloor$, where $\lfloor
	\frac{N}{2}\rfloor$ gives the integer part of $\frac{N}{2}$ (see Appendix C).
	
	Hence, most of the steady-state degeneracy is lifted once the Raman-assisted hopping is switched on, thanks to the competition between the cavity-induced gauge coupling and the Raman-assisted hopping.
	As we show below, this would give rise to hosts of quasi-steady eigenmodes and unique long-time dynamics under OBC.
	
	\section{Long-time dynamics and quasi-steady modes}

		Under the mean-field approximation, the cavity field is given by the expression
		$|\alpha|^2=\lambda^2\langle\hat{K}\rangle^2/(\Delta_c^2+\kappa^2).
	$ In Fig.~\ref{Fig3}(a)(b), we show the long-time damping behavior of the cavity field under OBC.
	The cavity field exhibits different exponential damping rates at different times, with a smaller rate at longer times.
	Note that, since the operator commutation relations are highly boundary-dependent, the cavity field remains constant under PBC (see Appendix E).
	Further, regardless of the initial state, the cavity damping is accompanied by
	an increased localization of the time-evolved state toward the boundary [see Fig.~\ref{Fig3}(c)(d)].
	
	The stage-wise slowdown in the steady-state-approaching dynamics can be understood by adopting a perturbative analysis under small $s$.
	We rewrite the Liouvillian as $\mathcal{L}=\mathcal{L}_0+s\mathcal{L}_1$, collecting all $s$-dependent terms in $s\mathcal{L}_1$, with
	$\mathcal{L}_1[\rho]=-i[\hat{K}+\hat{K}^\dag,\rho]$. We expand
	the density matrix as $\rho=\rho_s+\sum_{n=1} s^n \rho_n$, where $\rho_s$ is the steady state for $s=0$, with $\mathcal{L}_0[\rho_s]=0$.
	Formally, the steady state can be written as $\rho_s=\sum_{i,j}\alpha_{ij}|D^{B_i}_L\rangle \langle D^{B_j}_L|$, where $|D^{B_i}\rangle$ is a dark state of $\hat{K}$ in the subspace $\mathcal{H}_{B_i}$.
	We then define
	$B_m=\max(B_i,B_j)$, which characterizes the extent of spatial localization of the state $\rho_s$.
	
	Matching coefficients in $s$ order-by-order in the eigen equation $\mathcal{L}[\rho]=\lambda\rho$,
	one concludes that the density-matrix expansion should have infinite terms when $\rho$ is a steady state, with $\rho_n$ satisfying
	$\mathcal{L}_0[\rho_{n+1}]=-\mathcal{L}_1[\rho_n]$.
	For quasi-steady eigenmodes, the expansion would first truncate at some order $M$, with $\rho=\rho_s+\sum_{n=1}^{M} s^n \rho_n$.
	We derive an analytic expression of $\rho_n$ for the quasi-steady states (see Appendix D), revealing a power-law-scaling of their eigenvalues
	\begin{align}
		\lambda\sim s^{M+n_{L}+1}\operatorname{Tr} (\rho_{n_{L}} \mathcal{L}_1[\rho_{M}]),
	\end{align}
	where $n_{L}$ is the lowest order at which $\operatorname{Tr} (\rho_{n_{L}} \mathcal{L}_1[\rho_{M}]) \neq 0$.
	A lower bound of the exponent $p=M+n_{L}+1$ can also be derived, which monotonically increases with decreasing $B_m$ [see Fig.~\ref{Fig4}],
	connecting the eigenvalues of the quasi-steady states with localization.

	In Fig.~\ref{Fig4}(a)(c), we show the Liouvillian spectra $\text{Re}\lambda$ of several groups of quasi-steady modes,
	each group exhibiting similar power-law scaling with respect to $s$. As a consequence of the hierarchy of structures in the Liouvillian spectrum,
	under any given $s$, different groups of eigenmodes take turns to dominate the long-time dynamics.
	For instance, under $s/\gamma=0.5$, the eigenmodes marked by green square and triangle sequentially dominate the damping dynamics at typical times $t_q=1/\text{Re}\lambda$ [marked in Fig.~\ref{Fig3}(a)(b) by the corresponding symbols].
	The exponential damping rates of the cavity field, fitted using solid-lines in Fig.~\ref{Fig3}(a)(b), agree well with $\text{Re}\lambda$ at the corresponding markers.
	
	Importantly, different groups of eigenmodes, characterized by different exponents $p$, have distinct localization properties.
	In Fig.~\ref{Fig4}(b)(d), we plot the relation between the power-law exponent $p$ and $B_m$.
	The analytically obtained lower bounds lie below but close to the numerically calculated exponent.
	Since eigenmodes with $\text{Re}\lambda\sim s^p$ should dominate the dynamics at the time scale $t_q \sim s^{-p}$, for any given small hopping rate $s$,
	eigenmodes with larger $p$ (hence more localized) should dominate at longer times.
Thus, the OBC significantly enriches the long-time dynamics, exciting eigenmodes with localization-dependent eigenvalues out of the steady-state subspace, such that
more localized eigenmodes dominate the dynamics at later times.

	\section{Conclusion}
	We demonstrate the impact of NHSE in the many-body dynamics of an atom-cavity hybrid system.
	Besides inducing boundary localization of the steady states,
	the NHSE exhibits intriguing features that are absent in the single-particle case.
In particular, the NHSE gives rise to a stage-wise slowdown of the long-time dynamics under OBC.
The situation is completely contrary to the single-body case, where
the long-time dynamics is slower under the PBC, due to the finite Liouvillian gap under the OBC. Thus, the slowdown under OBC in the many-body case is
a direct manifestation of the NHSE beyond the single-body paradigm. 
Physically, this stage-wise slowdown indicates the presence of distinct groups of
quasi-steady states, with each group featuring a unique power-law-scaled Liouvillian eigenvalues, as well as a similar extent of boundary localization. 
Our results thus establish a general connection between spectral geometry and open-system dynamics, and reveal 
the rich spectral and dynamic consequences of non-Hermitian physics in many-body quantum open systems.

Given the recent progress in atom-cavity hybrid systems, our predictions can be experimentally checked either in real-space or momentum-space lattices~\cite{WuHaiBing2021,esslinger22,Esslinger@2023}. For instance, in the latter case, our model can be implemented in the momentum space of the atoms, following the design in Ref.~\cite{Esslinger@2023}. Therein, discrete momentum states of atoms in an optical cavity are coupled by Bragg processes consisting of pump lasers and the cavity field. While the discrete momentum states can be mapped to discrete lattice sites, the Bragg processes play the role of Raman-assisted hopping discussed in this work.
Further, the directional transfer along the momentum lattice is facilitated by the photon recoil of the Bragg processes (together with the cavity dissipation), such that tilting of the lattice as shown in Fig.~\ref{Fig1}(a) is no longer necessary.
Finally, the long-time dynamics can be probed by detecting the time evolution of the cavity field as well as the momentum-space distribution of atoms through the time-of-flight imaging.

	
	\acknowledgments{This research is supported by the National Natural Science Foundation of China (Grant No. 11974331), and by the Innovation Program for Quantum Science and Technology (Grant Nos. 2021ZD0301200, 2021ZD0301904, 2021ZD0302000).
		
		\renewcommand{\thesection}{\Alph{section}}
		\renewcommand{\theequation}{A\arabic{equation}}
		\setcounter{equation}{0}
		\section*{Appendix A: Steady-state degeneracy under the periodic boundary condition}
		
		Under the periodic boundary condition (PBC), $[\hat{K},\hat{K}^\dagger]=0$, therefore $\hat{K}$ and $\hat{K}^\dagger$ have common eigenstates and complex-conjugate eigenvalues. We denote their common eigenstates as $\{|\phi_{n\beta}\rangle\}$, where
		$n$ and $\beta$ respectively label the eigenvalue and the corresponding degenerate eigenstate.
		It follows that $\hat{K}|\phi_{n\beta}\rangle=\mu_n|\phi_{n\beta}\rangle$ and $\hat{K}^\dagger|\phi_{n\beta}\rangle=\mu_n^*|\phi_{n\beta}\rangle$, where
		$\mu_n$ is the $n$th eigenvalue with a degeneracy $\beta_n$.
		
		For any density matrix $\rho_s=\sum_{\beta,\beta'}\alpha_{\beta,\beta'}|\phi_{n\beta}\rangle\langle \phi_{n\beta'}|$ with superposition coefficients $\alpha_{\beta,\beta'}$, we have $[\hat{K},\rho_s]=[\hat{K}^\dagger,\rho_s]=0$. With direct calculation, we find
		\begin{equation}
			\begin{aligned}
				\mathcal{L}[\rho_s]&=-i[\hat{H},\rho_s]+\gamma(2\hat{K}\rho_s\hat{K}^\dagger-\{\hat{K}^\dagger\hat{K},\rho_s\})\\&=\gamma(2\mu_n\mu_n^*-2\mu_n^*\mu_n)\rho_s=0.
			\end{aligned}
		\end{equation}
		Therefore $\rho_s$ is a steady state under PBC. Taking $\rho_s$ at different $n$ and $\alpha_{\beta,\beta'}$, we can construct a total number of $\sum_n\beta_n^2$ independent steady states. We have numerically checked that, for all cases considered in our work, this number actually gives the exact steady-state degeneracy.
		Since $\beta_n\geq 1$, the steady-state degeneracy $\mathcal{D}=\sum_n\beta_n^2\geq\sum_n\beta_n=\frac{L!}{N!(L-N)!}$, where the latter is the dimension of the Hilbert space with $L$ sites and $N$ spinless fermions.
		\renewcommand{\thesection}{\Alph{section}}
		\renewcommand{\theequation}{B\arabic{equation}}
		\setcounter{equation}{0}
		\section*{Appendix B: Steady-state degeneracy at $s=0$ under the open boundary condition}
		
		We consider a one-dimensional lattice with $L$ sites and $N$ spinless fermions, and adopt the anti-symmetrized Fock basis $\{|j_1,j_2,\cdots, j_N\rangle\}$, where $j_i\in [1,L]$ indicates the site occupied by the $i$~th fermion. Here the lattice sites are labeled in ascending order from left to right.
		The Hilbert space is then divided into different subspaces $\mathcal{H}_B$, each with a fixed sum of the occupied-site labels $B=\sum^N_{i=1} j_i$.
		Here $B$ can only take integer values within the range of $B_{\text{min}}=N(N+1)/2\leq B\leq B_{\text{max}}=N(2L-N+1)/2$.
		It is straightforward to show that the quantum jump operator $\hat{K}$ transforms a state in the subspace $\mathcal{H}_B$ to one in $\mathcal{H}_{B-1}$. The matrix of $\hat{K}$ can therefore be written in a  block-off-diagonal form. We denote the off-diagonal block between the subspaces $\mathcal{H}_B$ and $\mathcal{H}_{B-1}$ using the transfer matrix $M_B$.
		Here $M_B$ is a $G_B\times G_{B-1}$ matrix, where $G_B$ is the dimension of the subspace $\mathcal{H}_B$. Formally, $G_B$ can be determined from the recursive relation
		\begin{align}
			G_B(L,N)&=\sum_{n=\lfloor B/N\rfloor+1}^L G_{B-n}(n-1,N-1),
			\label{recursive}
		\end{align}
		and the condition
		\begin{equation}
			G_B(L,2)=\text{max}\left[0,\text{min}(\lfloor \frac{B+1}{2}\rfloor-1, L-\lfloor \frac{B}{2}\rfloor)\right].
		\end{equation}
		Here $G_B(L,N)$ is the dimension of the corresponding subspace (fixed by $B$) for a lattice with $L$ sites and $N$ fermions, and $\lfloor \cdots \rfloor$ gives the  integer part of $\cdots$.
		
		We now examine the dark state $|D_L\rangle$ of $\hat{K}$, satisfying $\hat{K}|D_L\rangle=0$. Since the transfer matrix $M_B$ is in general not square, there are at least $\text{max}(G_{B-1}-G_{B},0)$ vectors (denoted as $v_B$) that satisfy $M_B v_B=0$, each corresponding to a dark state in the subspace of $\mathcal{H}_B$.
		Notice that the number of dark states (in the subspace of $\mathcal{H}_B$) is at the minimum when the rank of the matrix $M_B$ is full.
		On the other hand, for each given $L$ and $N$, $G_B$ monotonically increases with $B$ to a maximum value $G_m$, after which it decreases monotonically with $B$ [see Fig.~\ref{FigB1}(a)].
		It follows that, when all the transfer matrices between adjacent subspaces are considered, the number of dark states is at least $G_m$. Given the form of the Lindblad equation, the steady states, depicted by density matrices, have a degeneracy of at least $\mathcal{D}=G_m^2$ at $s=0$. The minimum degeneracy occurs only if all $M_B$ matrices are of full rank. Numerically, we have checked that this is true for all cases considered throughout the work.
		
		\begin{figure}[tbp]
			\includegraphics[width=0.45\textwidth]{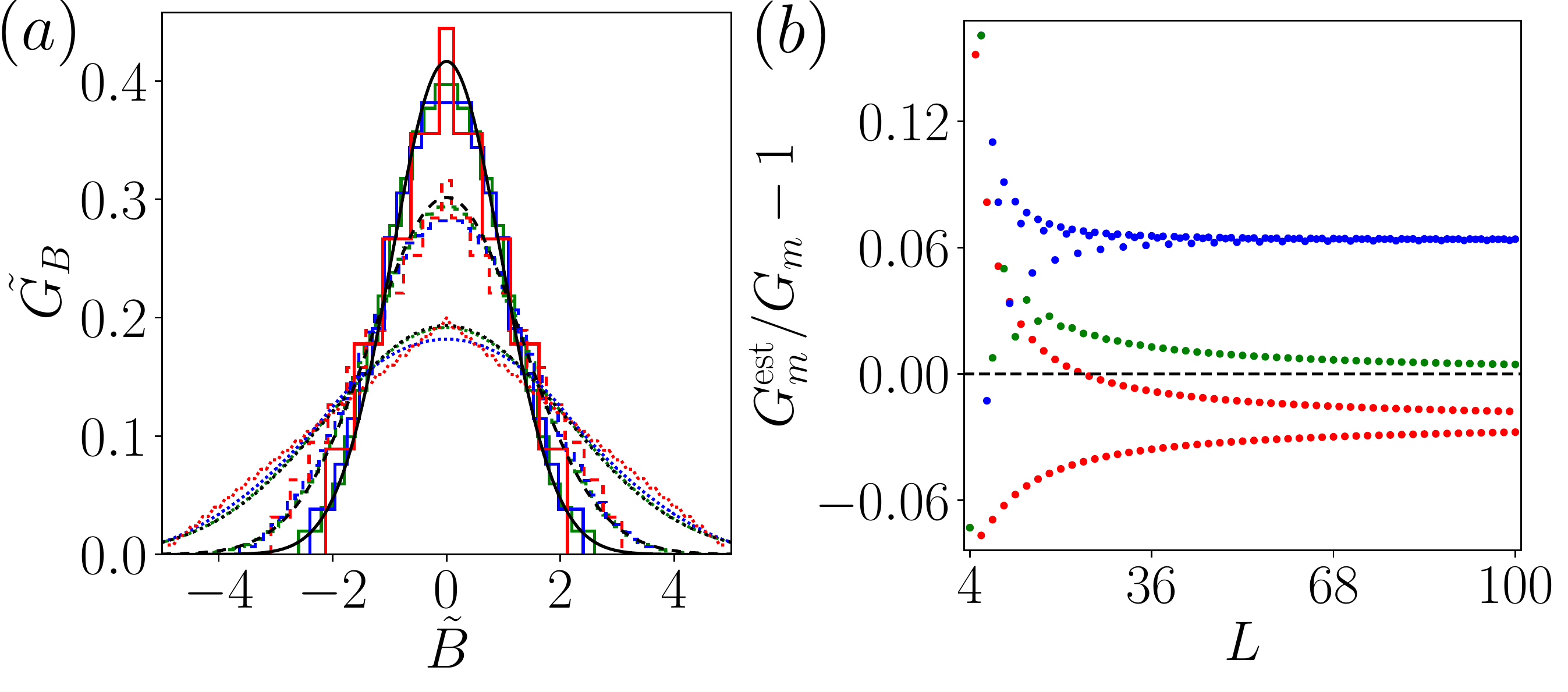}
			\caption{(a) The rescaled distribution function $\tilde{G}_B=\frac{\sqrt{N(L-N)}N!(L-N)!}{L!}G_B$ versus $\tilde{B}=\frac{B-N(L+1)/2}{\sqrt{N(L-N)}}$. The red, blue, and green lines are respectively calculated using $N=2$, $N=3$, and $N=L/2$. The black lines indicate the Gaussian distribution $\tilde{G}_B=\exp(-\tilde{B}^2/2\sigma^2)/\sqrt{2\pi}\sigma$, with the variance $\sigma^2=(L+1)/12$. The lattice size $L=10,20,50$ for solid, dashed and dotted lines. (b) Relative error between $G_m$ and its estimation $G^{\text{est}}$ using Eq.~(\ref{eq:suppGm}). The red, blue, and green dots are respectively $N=2,3$ and $N=L/2$ (with even $L$).
			} \label{FigB1}
		\end{figure}
		
		The exact value of $G_m$ is determined by the maximum of $G_B$ in the recursive relation Eq.~(\ref{recursive}). However, in the limit $L, N\gg 1$,
		we can estimate its value by adopting a statistical approach.
		More explicitly, we treat the atomic positions as dependent integer random variables uniformly distributed in the range of $[1,L]$, satisfying the Pauli exclusion principle. We can then view $G_B/(\frac{L!}{N!(L-N)!})$ as the distribution function of $B=\sum_i j_i$. The mean and variance of the position for the $i$th atom are respectively $E[j_i]=\frac{L+1}{2}$ and $\sigma^2[j_i]=\frac{(L+1)(L-1)}{12}$. The covariance of the position for any two atoms $i$ and $k$ is $\text{Cov}[j_i,j_k]=-\frac{L+1}{12}$.
		From these, we calculate the mean and variance for $B=\sum_ij_i$, as
		\begin{align}
			E[B]&=\sum_iE[j_i]=\frac{N(L+1)}{2},\\  \sigma^2[B]&=\sum_i\sigma^2[j_i]+\sum_{i\neq k}\text{Cov}[j_i,j_k]=\frac{N(L-N)(L+1)}{12},
		\end{align}
		and the correlation coefficient
		\begin{align}
			\eta[j_i,\sum_{k\neq i}j_k]=\frac{\text{Cov}[j_i,\sum_{k\neq i}j_k]}{\sqrt{\sigma^2[j_i]\sigma^2\sum_{k\neq i}[j_k]}}=-\sqrt{\frac{N-1}{(L-N+1)(L-1)}}.
		\end{align}
		Since $\eta[j_i,\sum_{k\neq i}j_k]$ converges to $0$ with increasing $L$, we apply a dependent central limit theorem~\cite{DCLT}, and find that the distribution of $B$ converge to a discrete Gaussian distribution under $L,N\gg 1$.
		
		In Fig.~\ref{FigB1}(a), we plot a rescaled $G_B$ for different values of $L$ and $N$. With a proper rescaling, $G_B$ is very close to the Gaussian distribution even for $N=2,3$, as considered in this work.
		
		Since the Hilbert-space dimension is $\frac{L!}{N!(L-N)!}$, the maximum dimension of the subspace with a fixed $B$ is approximately
		\begin{equation}
			\begin{aligned}
				G_m&\approx G^{\text{est}}_m=\frac{1}{\sqrt{2\pi\sigma^2[B]}}\frac{L!}{N!(L-N)!}\\&=\sqrt{\frac{6}{\pi N(L-N)(L+1)}}\frac{L!}{N!(L-N)!}.
			\end{aligned}
			\label{eq:suppGm}
		\end{equation}
		In Fig.~\ref{FigB1}(b), we plot the relative error between the exact value of $G_m$ and its estimation using Eq.~(\ref{eq:suppGm}). While the error approaches 0 with increasing $N$ and $L$, it is already smaller than $10\%$ for $N=2,3$.

		Finally, since the matrix of $\hat{K}$ is block-off-diagonal, any dark state $|D_L\rangle$ can be written as $|D_L\rangle=\sum_B\alpha_B|D_L^B\rangle$,
		where $|D_L^B\rangle$ is the dark state in the subspace $\mathcal{H}_{B}$, and $\alpha_B$ is the weight of the superposition.
		For the convenience of later discussions, we note that, since $P\hat{K}P^{-1}=\hat{K}^\dag$ (where $P$ is the spatial-inversion operator), $|D_R^{B_{\text{max}}-B}\rangle=P|D_L^B\rangle$ is a dark state of $\hat{K}^\dag$ in the subspace $\mathcal{H}_{B_{\text{max}}-B}$, with $\hat{K}^\dag|D_R^{B_{\text{max}}-B}\rangle=0$.
		\renewcommand{\thesection}{\Alph{section}}
		\renewcommand{\theequation}{C\arabic{equation}}
		\setcounter{equation}{0}
		\section*{Appendix C: Steady states under finite $s,\gamma$ with OBC}		
		
		To find the steady states under finite $s,\gamma$, we start by considering the simplest case of $\gamma=0$. When $\gamma=0$, the system is no longer an open system, and the steady states are generated by the eigenstates of the Hermitian Hamiltonian. Assuming that the Hamiltonian $H$ has $a_n$ eigenstates with $n$-fold degeneracy, the steady-state degeneracy is then $\sum_n a_n n^2$, which is typically a large number.
		
		However, with a very small $\gamma$, most of these states become perturbed and are no longer steady state of the system.
		Their Liouvillian eigenvalues exhibit linear dependence with $\gamma$.
		The remaining steady states can be formally written as
		$\rho_s=\sum_{n=0}\gamma^n \rho_n$, where
		\begin{align}
			\gamma^n\rho_n=\sum_{q=0}^{n}A^{-q}B^{-(n-q)}\hat{K}^q(\hat{K}^\dagger)^{n-q}\rho_0.
			\label{eq:gamnrhon}
		\end{align}
		Here $A=\frac{-is}{(1+i\Delta_c/\kappa)\gamma}$, $B=\frac{is}{(1-i\Delta_c/\kappa)\gamma}$ are the same as those in Eq.~(\ref{eq:rhonform}), and the generator $\rho_0$ satisfy $[\hat{K},\rho_0]=[\hat{K}^\dagger,\rho_0]=0$
		
		Equation (\ref{eq:gamnrhon}) can be derived by induction.
		For this purpose, we write the Liouvillian as
		$\mathcal{L}=\mathcal{L}_0'+\gamma\mathcal{L}_1'$, where
		\begin{align}
			\mathcal{L}'_0&=-is[\hat{H}_1,\rho],\\
			\mathcal{L}'_1[\rho]&=-\frac{i}{\gamma}[\hat{H}_0,\rho]+\left(2\hat{K}\rho\hat{K}^\dagger-\left\{\hat{K}^\dagger\hat{K},\rho\right\}\right),
		\end{align}
		and
		\begin{align}
			\hat{H}_0&=-\frac{\Delta_c}{\kappa}\gamma\hat{K}^\dagger\hat{K}, \label{eq:suppH0}\\
			\hat{H}_1&=\hat{K}+\hat{K}^\dagger. \label{eq:suppH1}
		\end{align}
		The steady-state condition $\mathcal{L}[\rho_s]=0$ requires
		\begin{align}
			\mathcal{L}[\rho_0]&=\gamma \mathcal{L}_1'[\rho_0]\\
			\mathcal{L}[\rho_n]&=-\mathcal{L}_1'[\rho_{n-1}]+\gamma\mathcal{L}_1'[\rho_n], \quad \text{for  } n>1  .\label{eq:sind2}
		\end{align}
		
		At the zeroth order, $\rho_0$ satisfies Eq.~(\ref{eq:gamnrhon}). Now if up to the $n$th order, $\rho_n$ satisfies Eq.~(\ref{eq:gamnrhon}), using Eq.~(\ref{eq:sind2}), we have
		\begin{equation}
			\begin{aligned}
				\gamma \mathcal{L}&_1'[\gamma^n\rho_n]=2\gamma\sum_{q=0}^{n}A^{-q}B^{-(n-q)}\hat{K}^{q+1}(\hat{K}^\dagger)^{n-q+1}\rho_0\\&-(1+\frac{\Delta_c}{\kappa}i)\gamma\sum_{q=0}^{n}A^{-q}B^{-(n-q)}\hat{K}^\dagger \hat{K}^{q+1}(\hat{K}^\dagger)^{n-q}\rho_0\\
				&-(1-\frac{\Delta_c}{\kappa}i)\gamma\sum_{q=0}^{n}A^{-q}B^{-(n-q)} \hat{K}^{q}(\hat{K}^\dagger)^{n-q+1}\hat{K}\rho_0\\
				=&-\mathcal{L}_0'[\gamma^{n+1}\rho_{n+1}].
			\end{aligned}
		\end{equation}
		The expression for $\rho_{n+1}$ can then be solved, and is found to be consistent with Eq.~(\ref{eq:gamnrhon}).
		
		Notice that $\hat{K}$ transforms a state in the subspace $\mathcal{H}_B$ to one in $\mathcal{H}_{B-1}$. Combined with the
		condition $B\in [B_{\text{min}},B_{\text{max}}]$, we have  $\hat{K}^{q}=(\hat{K}^\dagger)^{q}=0$ if $q> B_{\text{max}}-B_{\text{min}}=NL$.
		It follows that there are only finite terms in the expansion $\rho_s=\sum_{n=0}\gamma^n \rho_n$, yielding the analytic form of the
		steady state.
		
		With Eq~(\ref{eq:gamnrhon}), each distinct $\rho_0$, with $[\hat{K},\rho_0]=[\hat{K}^\dagger,\rho_0]=0$, would give a distinct steady state.
		In the following, we count the steady-state degeneracy by constructing different $\rho_0$.
		We start by considering the following dark states of $\hat{K}$
		\begin{align}
			|D_L^{B(m)}\rangle=|1,2,...N-2m\rangle\otimes|C^{2m}_{L-2N+4m}(N-2m)\rangle,
		\end{align}
		where $|C^{N}_L(d)\rangle$ is recursively defined for even $N,L$ as
		
		\begin{widetext}
			\begin{equation}
				|C^{N}_L(d)\rangle=\left\{
				\begin{array}{lr}
					\frac{1}{\sqrt{L/2}}\sum_{q=1}^{L/2}(-1)^{q}|q+d\rangle\otimes |L+1-q+d\rangle & \text{for} \quad N=2\\
					\frac{1}{\sqrt{(L-N)/2+1}}\sum_{q=1}^{(L-N)/2+1}(-1)^{qN/2}|q+d\rangle\otimes |C_{L-2q}^{N-2}(q+d)\rangle\otimes|L+1-q+d\rangle & \text{for} \quad N\geq 4 \\
				\end{array}
				\right.
				,
			\end{equation}
		\end{widetext}
		Specifically, $|D_L^{B(m)}\rangle$ is a dark state of $\hat{K}$ in the subspace $\mathcal{H}_{B(m)}$ with $B(m)=N(N+1)/2+m(L-2N+2m)$.
		Note that, $|D_R^{N(L+1)-B(m)}\rangle=P|D_L^{B(m)}\rangle$ is a dark state of $\hat{K}^\dagger$, which is also the spatial-inversion of $|D_L^{B(m)}\rangle$.
		
		We further define a series of states $|\varphi_{i,j}\rangle$, such that
		\begin{equation}
			\begin{aligned}
				\hat{K}|\varphi_{i,j_i}\rangle&=\sum_{j=1}^{n_{i-1}}|\varphi_{i-1,j}\rangle\langle \varphi_{i-1,j}|\hat{K}|\varphi_{i,j_i}\rangle\\&=\sum_{k,j}^{}|\varphi_{k,j}\rangle\langle \varphi_{k,j}|\hat{K}|\varphi_{i,j_i}\rangle,
				\label{eq:Kphiij}
			\end{aligned}
		\end{equation}
		\begin{equation}
			\begin{aligned} \hat{K}^\dagger|\varphi_{i,j_i}\rangle&=\sum_{j=1}^{n_{i+1}}|\varphi_{i+1,j}\rangle\langle \varphi_{i+1,j}|\hat{K}^\dagger|\varphi_{i,j_i}\rangle\\&=\sum_{k,j}^{}|\varphi_{k,j}\rangle\langle \varphi_{k,j}|\hat{K}^\dagger|\varphi_{i,j_i}\rangle,
				\label{eq:Kphidagij}
			\end{aligned}
		\end{equation}
		with $|\varphi_{B(m),1}\rangle=|D_L^{B(m)}\rangle$, and $|\varphi_{N(L+1)-B(m),1}\rangle=|D_R^{N(L+1)-B(m)}\rangle$. In deriving Eqs.~(\ref{eq:Kphiij}) and (\ref{eq:Kphidagij}), we used the property that $\hat{K}$ and $\hat{K}^\dagger$ only couple states with adjacent $B$ labels.
		
		For $\rho_0=\sum_{i,j_i}|\varphi_{i,j_i}\rangle\langle\varphi_{i,j_i}|$, with Eqs.~(\ref{eq:Kphiij}) and (\ref{eq:Kphidagij}), we find $\hat{K}\rho_0=\rho_0\hat{K}=\rho_0\hat{K}\rho_0$ and $\hat{K}^\dagger\rho_0=\rho_0\hat{K}^\dagger=\rho_0\hat{K}^\dagger\rho_0$. The total number of steady states corresponds to the total number of distinct $|D_L^{B(m)}\rangle$, which is given by
		\begin{align}
			\mathcal{D}=1+\frac{1+(-1)^L}{2}\lfloor \frac{N}{2}\rfloor.
		\end{align}
		We have numerically checked that this is exactly the steady-state degeneracy at finite $s,\gamma$ for all cases considered in this work.
		\renewcommand\arraystretch{1.5}
		\begin{table*}
			\setlength{\tabcolsep}{2mm}
			\begin{tabular}{|c|c|c|c|}		
				\hline
				$\mathcal{D}$ &PBC&OBC, $s=0$&OBC, finite $s/\gamma$\\
				\hline
				Exact degeneracy&$\sum_n \beta_n^2$&$G_m^2$&$1+\frac{1+(-1)^L}{2}\lfloor \frac{N}{2}\rfloor$\\
				\hline
				Estimation value&$\geq \frac{L!}{N!(L-N)!}$&$\approx \frac{6}{\pi N(L-N)(L+1)}(\frac{L!}{N!(L-N)!})^2$&\\
				\hline
			\end{tabular}
			\caption{Exact value and its estimation of the steady state degeneracy under different conditions.}
			\label{tableA1}
		\end{table*}
		
		\begin{figure*}[tbp]
			\includegraphics[width=0.7\textwidth]{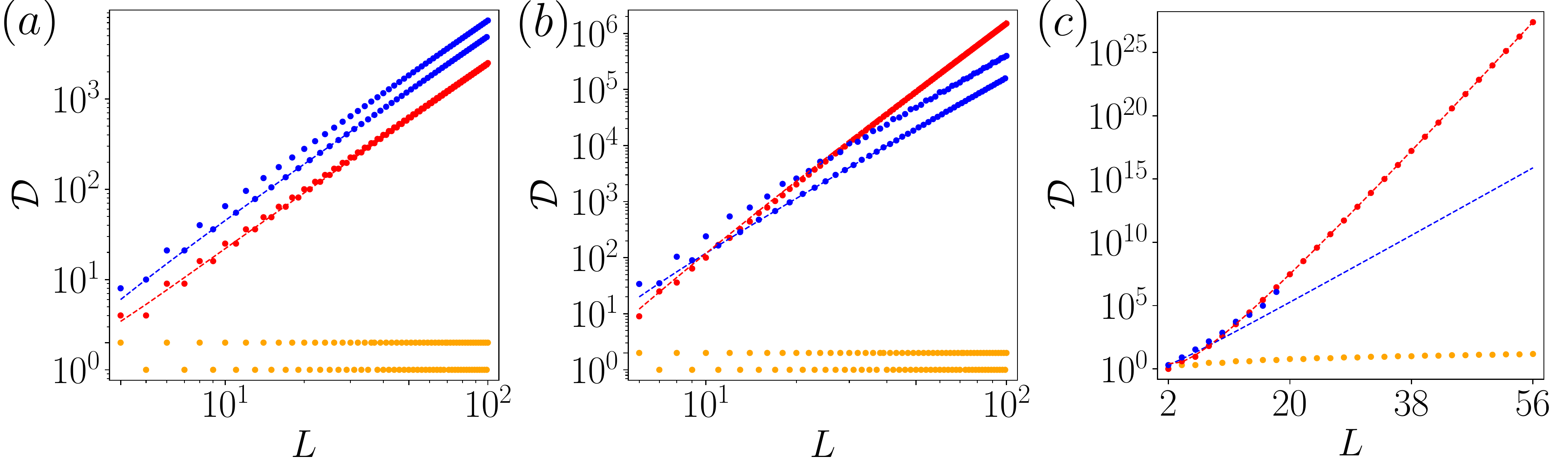}
			\caption{Steady-state degeneracies under PBC (blue), OBC with $s=0$ (red) and finite $s/\gamma$ (orange), respectively. The blue dashed lines indicate the dimension of the Hilbert space $\frac{L!}{N!(L-N)!}$. The red dashed lines indicate the estimated values of $G_m^2$. The fermion numbers are respectively (a) $N=2$, (b) $N=3$, (c) $N=L/2$ (with even $L$).
			} \label{FigC1}
		\end{figure*}
		
		Now that we have derived the steady-state degeneracy under different conditions, we summarize the results in Table~\ref{tableA1} and Fig.~\ref{FigC1}. With a fixed atom number $N$, the degeneracy under PBC or OBC at $s=0$ has a power-law relationship with the lattice size $L$ [see Fig.~\ref{FigC1}(a)(b)].  With a fixed filling, the degeneracy under PBC or OBC at $s=0$ has an exponential relationship with $L$ [see Fig.~\ref{FigC1}(c)]. For cases considered in this work (with $N=2,3$ and $L\sim 10$), the degeneracy is the largest under PBC, which is reduced significantly under OBC, and further decreased (also significantly) at a finite $s$.
		\renewcommand{\thesection}{\Alph{section}}
		\renewcommand{\theequation}{D\arabic{equation}}
		\setcounter{equation}{0}
		\section*{Appendix D: Perturbative characterization of the Liouvillian spectrum}
		Under finite $s$, some (or most) degenerate steady states for $s=0$ become perturbed, and are no longer steady states.
		The Liouvillian gap of the system (under OBC) is determined by these states, which have a dramatic impact on the long-time dynamics as we show in the main text.
		
		From numerical simulations, we find that $\text{Re}\lambda$, the real components of the Liouvillian eigenvalues, of these perturbed states manifest distinct power-law scalings with $s$. In this section, we show what these perturbed states are, and what the exponents of the power-law scalings are.
		
		We start by devising a perturbation series for the density matrix,
		by writing the Liouvillian as $\mathcal{L}=\mathcal{L}_0+s\mathcal{L}_1$. Here
		\begin{align}
			\mathcal{L}_0[\rho]&=-i[\hat{H}_0,\rho]+\gamma\left(2\hat{K}\rho\hat{K}^\dagger-\left\{\hat{K}^\dagger\hat{K},\rho\right\}\right),\\
			\mathcal{L}_1[\rho]&=-i[\hat{H}_1,\rho],
		\end{align}
		where $H_0$ and $H_1$ are defined in Eq.~(\ref{eq:suppH0}) and (\ref{eq:suppH1}), respectively.
		
		The steady-state density matrices $\rho_s$ satisfy $\mathcal{L}_0[\rho_s]=0$. Motivated by the power-law dependence of $\text{Re}\lambda$ on the parameter $s$, we make the following expansion of the density matrix $\rho=\rho_s+\sum_{n=1} s^n \rho_n$.
		For a steady state, $\mathcal{L} [\rho]=0$ should hold at each given order of $s$, we then have $\mathcal{L} _0[\rho_{n}]=-\mathcal{L}_1[\rho_{n-1}]$. This leads to the series expansion
		\begin{align}
			\mathcal{L} [\rho_s]&=s\mathcal{L}_1[\rho_s],\\
			\mathcal{L} [\rho_1]&=-\mathcal{L}_1[\rho_s]+s\mathcal{L}_1[\rho_1]\\
			\mathcal{L} [\rho_2]&=-\mathcal{L}_1[\rho_1]+s\mathcal{L}_1[\rho_2],\\
			&\cdots\\
			\mathcal{L} [\rho_n]&=-\mathcal{L}_1[\rho_{n-1}]+s\mathcal{L}_1[\rho_n],
		\end{align}
		which should continue indefinitely for a steady state.
		
		For a quasi-steady mode, the series must terminate at some order $M$, where one cannot find any $\rho_{M+1}$ to satisfy $\mathcal{L}_0 [\rho_{M+1}]= -\mathcal{L}_1[\rho_{M}]$. Keeping the density matrix expansion up to the $M$th order $\rho=\rho_s+\sum_{n=1}^{M} s^n \rho_n$, we have
		\begin{align}
			\mathcal{L}[\rho]=s^{M+1}\mathcal{L}_1[\rho_{M}].
		\end{align}
		Perturbatively, the leading order of the eigenvalue $\lambda$ for this quasi-steady mode is
		\begin{align}
			\lambda\sim s^{M+n_L+1}\operatorname{Tr} (\rho_{n_L} \mathcal{L}_1[\rho_{M}]),
			\label{eq:bound}
		\end{align}
		where $n_L$ is the lowest order at which $\operatorname{Tr} (\rho_{n_L} \mathcal{L}_1[\rho_{M}]) \neq 0$.
		Note that the conclusion above relies on the observation that, under small but finite $s$, $\text{Re}\lambda$ of the low-lying quasi-steady modes have power-law scalings.
		In the following, we will derive the lower bound of the exponent $p_{\text{min}}=M+n_L+1$, which characterizes the lowest-order exponent of the power-law scaling.
		In the process, we will also reveal the nature of the low-lying quasi-steady modes.
		
		Before proceeding, we prove the following theorem.\\
		
		\textbf{Theorem 1.} For a given operator $O$ and any state $|\phi\rangle$, there exists a state $|\psi\rangle$ such that $O|\psi\rangle=|\phi\rangle$, if and only if $\langle \phi|O_D\rangle=0$ holds for all $|O_D\rangle$. Here $|O_D\rangle$ is the dark state of $O^\dag$, with $O^\dag|O_D\rangle=0$.
		
		Proof: We first write the singular value decomposition of $O$ as $O=U\Sigma V^\dagger$, where $U$ and $V^\dagger$ are unitary matrices, and $\Sigma=\operatorname{diag}\{\Sigma_i\}$ is a real diagonal matrix with non-negative diagonal elements given by $\{\Sigma_i\}$. We then define a matrix $\tilde{\Sigma}=\operatorname{diag}\{\tilde{\Sigma}_i\}$, where $\tilde{\Sigma}_i=1/\Sigma_i$ for $\Sigma_i\neq 0$, and $\tilde{\Sigma}_i=0$ otherwise.
		
		On the one hand, notice that $O^\dagger U(I-\Sigma \tilde{\Sigma})U^\dagger|\phi\rangle=0$, so that $U(I-\Sigma \tilde{\Sigma})U^\dagger|\phi\rangle$ is a dark state of $O^\dag$. Under the condition that $\langle\phi|O_D\rangle=0$ for all $|O_D\rangle$, we must have $\langle \phi|U(1-\Sigma \tilde{\Sigma})U^\dagger|\phi\rangle=\langle \phi|U(1-\Sigma \tilde{\Sigma})^\dagger(1-\Sigma \tilde{\Sigma})U^\dagger|\phi\rangle= 0$. This leads to $\Sigma \tilde{\Sigma}U^\dagger|\phi\rangle=U^\dag|\phi\rangle$. We then construct
		$|\psi\rangle=V\tilde{\Sigma}U^\dagger|\phi\rangle$, such that $O|\psi\rangle=U\Sigma \tilde{\Sigma}U^\dagger|\phi\rangle=|\phi\rangle$. This shows the existence of $|\psi\rangle$.
		In the derivations above, we have used the relations $\Sigma=\Sigma\Sigma\tilde{\Sigma}$ and $(I-\Sigma \tilde{\Sigma})^\dagger(I-\Sigma \tilde{\Sigma})=(I-\Sigma \tilde{\Sigma})$, where $I$ is the identity matrix.
		
		On the other hand, if $O|\psi\rangle=|\phi\rangle$, we must also have $\langle\phi|O_D\rangle=\langle\psi|O^\dag|O_D\rangle=0$ for any $|O_D\rangle$. Theorem 1 is thus proved.\\
		
		As a direct consequence of Theorem 1, for any operator $O$ with $O|\psi\rangle=|\phi\rangle$, we can construct another operator $\tilde{O}=V\tilde{\Sigma}U^\dagger$, where $V$, $\Sigma$ and $U$ are defined through the singular value decomposition $O=U\Sigma V^\dagger$.
		Provided $\langle\phi|O_D\rangle=0$, we have $\tilde{O}|\phi\rangle=|\psi\rangle$, that is, the operator $\tilde{O}$ serves as the inverse of $O$ in the relevant subspace. Applying the statement above to the operators $\hat{K}$ and $\hat{K}^\dag$, we can define $\tilde{K}$ and $\tilde{K}^\dag$, despite $\hat{K}$ and $\hat{K}^\dag$ being singular and non-invertible in general.
		
		With this construction, if $\langle D_R|\tilde{K}^{q}\rho_s=0$  hold for any dark state $|D_R\rangle$ (assuming $q<n$), the closed-form expression of $\rho_n$  can be written as
		\begin{align}
			s^n\rho_n=\sum_{q=0}^{n}A^q B^{n-q}\tilde{K}^{q}\rho_s (\tilde{K}^{\dagger})^{(n-q)}.
			\label{eq:rhonform}
		\end{align}
		Here $A=\frac{-is}{(1+i\Delta_c/\kappa)\gamma}$, $B=\frac{is}{(1-i\Delta_c/\kappa)\gamma}$, $\rho_s$ is a steady state at $s=0$, satisfying $\hat{K}\rho_s=\rho_s\hat{K}^\dagger=0$.
		
		We now derive Eq.~(\ref{eq:rhonform}) by induction. With direct calculation,
		\begin{align}
			\mathcal{L}_1[\rho_s]=-is\hat{K}^\dagger\rho_s+is\rho_s\hat{K}.
		\end{align}
		Since $\mathcal{L}_0[s\rho_1]=-s\mathcal{L}_1[\rho_s]$, under the condition $\langle D_R|\tilde{K}\rho_s=0$ and $\rho_s\tilde{K}^{\dagger}|D_R\rangle=0$ for any $|D_R\rangle$, $\rho_1$ is formally given by
		\begin{align}
			s\rho_1=\frac{-is}{(1+i\Delta_c/\kappa)\gamma}\tilde{K}\rho_s+\frac{is}{(1-i\Delta_c/\kappa)\gamma}\rho_s \tilde{K}^{\dagger}.
			\label{eq:Srho1}
		\end{align}
		Here we have applied Theorem 1 to define $\tilde{K}$ and $\tilde{K}^\dag$. Equation (\ref{eq:Srho1}) is consistent with Eq.~(\ref{eq:rhonform}).
		
		Now if up to the $n$th order, $\rho_n$ has the expression of Eq.~(\ref{eq:rhonform}), then in the $(n+1)$~th order, we have
		\begin{equation}
			\begin{aligned}
				s\mathcal{L}_1[s^n \rho_n]=&is\sum_{q=0}^{n-1}A^q B^{n-q}\tilde{K}^{q}\rho_s (\tilde{K}^{\dagger})^{n-q-1}\\&-is\sum_{q=1}^{n}A^q B^{n-q}\tilde{K}^{(q-1)}\rho_s (\tilde{K}^{\dagger})^{n-q}\\
				&-is\sum_{q=0}^{n}A^q B^{n-q}\hat{K}^\dagger \tilde{K}^{q}\rho_s (\tilde{K}^{\dagger})^{n-q}\\&+is\sum_{q=0}^{n}A^q B^{n-q} \tilde{K}^{q}\rho_s (\tilde{K}^{\dagger})^{n-q}\hat{K}\\
				=&-\mathcal{L}_0[s^{n+1}\rho_{n+1}].
				\label{eq:ind1}
			\end{aligned}
		\end{equation}
		In the derivations above, we have assumed $\langle D_R|\tilde{K}^{n+1}\rho_s=0$, $\rho_s\tilde{K}^{\dagger n+1}|D_R\rangle=0$.
		The expression of $\rho_{n+1}$ can be solved from Eq.~(\ref{eq:ind1}), and is consistent with Eq.~(\ref{eq:rhonform}).
		We have therefore derived Eq.~(\ref{eq:rhonform}) by induction. \\
		
		Equipped with Eq.~(\ref{eq:rhonform}), we are now ready to analyze the power-law scaling of the quasi-stead states.
		We start from a general steady state under $s=0$, given by $\rho_s=\sum_{i,j}\alpha_{i,j}|D^{B_i}_L\rangle\langle D^{B_j}_{L}|$, where $\alpha_{i,j}$ are the superposition coefficients, and the superscript $B_i$ indicates the dark state $|D^{B_i}_{L}\rangle$ is in the subspace of $\mathcal{H}_{B_i}$. We define $B_m$ to be the largest of all superscripts $\{B_i,B_j\}$ with nonzero $\alpha_{i,j}$.
		
		We are interested in determining the least order $l$ at which $\langle D_R|\tilde{K}^{l}|D^{B_m}_{L}\rangle\neq 0$.
		Physically, $|D^{B}_R\rangle$ ($|D^{B}_L\rangle$) is localized to the right (left) side of the lattice, and the extent of the localization is indicated by $B$.
		For the overlap integral $\langle D_R|\tilde{K}^{l}|D^{B_i}_{L}\rangle$ to be nonzero, a larger $l$ is required when $B_i$ becomes smaller, that is, when $|D^{B_i}_{L}\rangle$ becomes more and more localized toward the left.
		Thus, assuming $l$ is the least order at which $\langle D_R|\tilde{K}^{l}|D^{B_m}_{L}\rangle\neq 0$, we must have
		$\langle D_R|\tilde{K}^{n}|D^{B_n}_{L}\rangle= 0$ for any $n<l$.
		This leads to $\langle D_R|\tilde{K}^{n}\rho_s=0$ and $\rho_s\tilde{K}^{\dagger n}|D_R\rangle= 0$ for any $n<l$.
		Following the derivation of Eq.~(\ref{eq:rhonform}), the perturbative expansions of the quasi-steady states above $\rho_s$ are given by $\rho_n$ ($n<l$) in the form of Eq.~(\ref{eq:rhonform}).
		
		More explicitly, there can be two scenarios.
		First, when $l>0$, the quasi-steady states are given by
		$\rho=\rho_s+\sum_{n=1}^{l}s^n \rho_n$, with $\mathcal{L} [\rho_n]=-\mathcal{L}_1[\rho_{n-1}]+s\mathcal{L}_1[\rho_n]$ for any $n\leq l$. According to Eq.~(7), we have $M=l$. And by matching the total occupied-site label $B$ on either side of Eq.~(8), we have $n_L=l-1$. This gives $p_{\text{min}}=2l$.
		
		Alternatively, when $l=0$, $\rho=\rho_s$, meaning $M=0$. Since $\operatorname{Tr}(\rho_s \mathcal{L}_1 [\rho_s]) =0$ (determined by counting the label $B$ of $\rho_s$ and $\mathcal{L}_1[\rho_s]$), we have $n_L\geq 1$ according to Eq.~(8). This gives $p_{\text{min}}=2$.
		
		To summarize the above, we have
		\begin{align}
			\text{Re}\lambda \sim s^p, \quad \text{where}\,
			\left\{
			\begin{array}{lc}
				p\geq 2l\quad &\text{for} \quad l>0 \\
				p=2\quad &\text{for} \quad l=0\\
			\end{array}
			\right. .
		\end{align}
		
		Finally, since the low-lying quasi-steady states feature different $p_{\text{min}}$, they reside in different subspaces labeled by distinct $B$, and hence have different localization characteristics. The stage-wise slowdown of the long-time dynamics is therefore due to the dominance of quasi-steady states with different localizations at different time scales.
		
		In Table~\ref{tableA2}, we summarize the real components of the Liouvillian eigenvalues at the marked times in Fig.~4(a)(b) and the numerically fitted damping rate, obtained by fitting the solid-line segments in Fig.~3(a)(b). Their excellent agreement directly confirms the stage-wise dominance of quasi-steady modes at the corresponding times.
		\renewcommand\arraystretch{1.2}
		\begin{table}[h]
			\setlength{\tabcolsep}{1mm}
			\begin{tabular}{|c|c|c|c|}		
				\hline
				markers& green  &blue &purple\\
				\hline
				$|\rm{Re}(\lambda)|$&0.125,.00108&0.0366,.00220&0.0280,.00107\\
				\hline
				-fitted exponent&0.255,.00108&0.0322, 0.00219&0.0262,.00106\\
				\hline
			\end{tabular}
			\caption{$|\rm{Re}(\lambda)|$ and fitted exponent near the markers (square, triangle) in Fig.~3(a)(b). }
			\label{tableA2}
		\end{table}
		\renewcommand{\thesection}{\Alph{section}}
		\renewcommand{\theequation}{E\arabic{equation}}
		\setcounter{equation}{0}
		\section*{Appendix E: Comparison of the Dynamics under PBC and OBC}
		
		In our system, the boundary condition dramatically modifies the dynamics. We here illustrate this by studying the dynamics of two typical observables, the cavity field $|\alpha|^2$ and the bulk current $\langle \hat{J}\rangle$.
		The bulk current operator is defined as
		\begin{align}
			\hat{J}=\sum_i\hat{J}_i=2\gamma\hat{K}^\dagger\hat{K}-is(\hat{K}-\hat{K}^\dagger),
		\end{align}
		where $\hat{J}_i$ satisfies $\partial_t\hat{n}_i+(\hat{J}_i-\hat{J}_{i-1})=0$, and $\hat{n}_i$ is the fermionic number operator on site $i$.
		
		Under PBC, since $[\hat{K},\hat{K}^\dagger]=0$, the operators $\hat{K}$ and $\hat{J}$ commute with both the Hamiltonian and the quantum jump operator. Thus, $|\alpha|^2$ and $\langle \hat{J}\rangle$ are both conserved quantities, determined only by the initial state [see the black lines in Fig.~\ref{FigE1}]. Particularly, a constant bulk flow emerges if the initial state has a non-zero current, which is consistent with the non-Hermitian skin effect at the single-particle level~\cite{fang20}.
		
		However, under OBC, the cavity field and the bulk current are not conserved, but approach their steady-state values in a stage-wise fashion. The steady-state value of the bulk current $\langle \hat{J}\rangle$ is always $0$, since all atoms accumulate to the boundary in the steady states [see Fig.~\ref{FigE1}(b)].
		
		\begin{figure}[tbp]
			\includegraphics[width=0.45\textwidth]{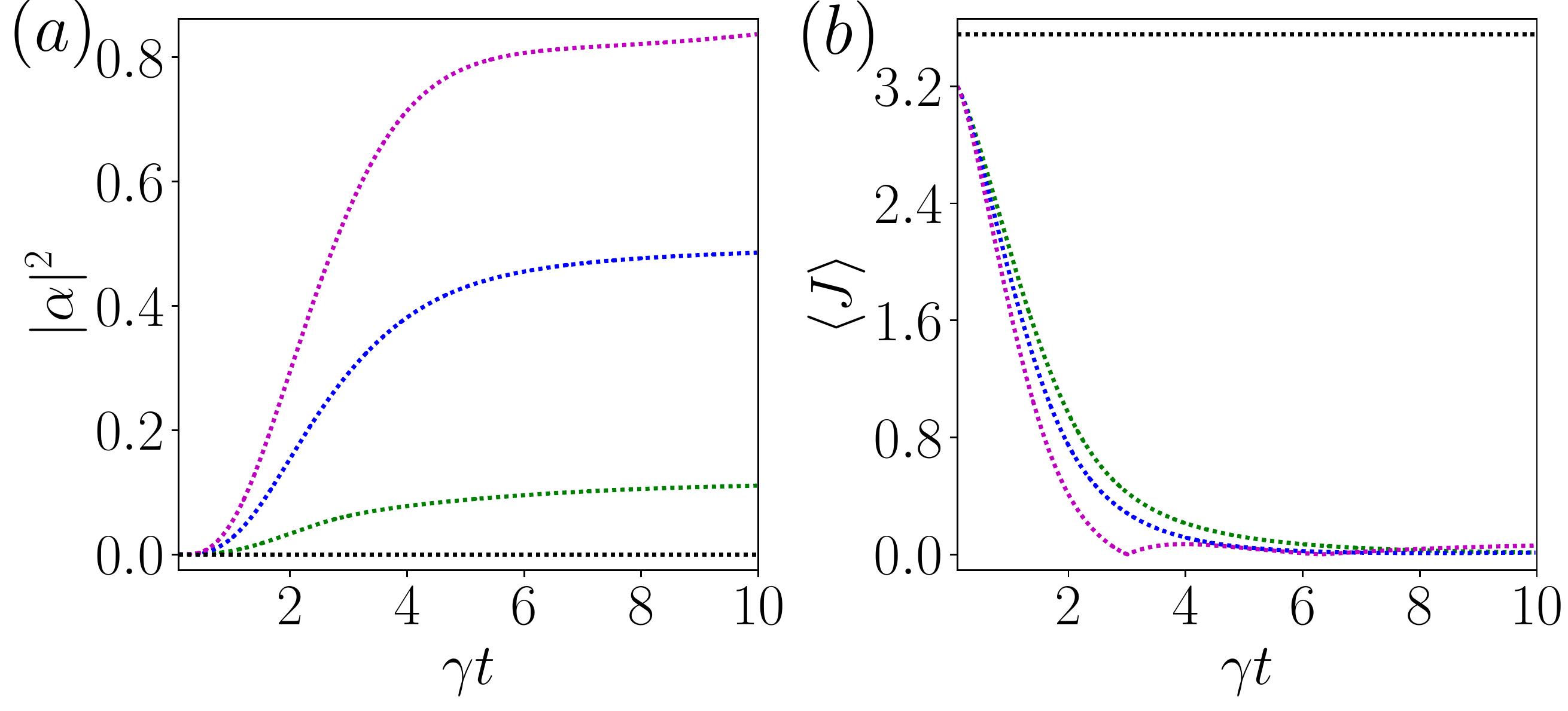}
			\caption{Dynamics of the cavity field $|\alpha|^2$ and the bulk current $\langle J\rangle$ under PBC (black, regardless of $s$), and under OBC with $s/\gamma=0.5,1.1,1.6$ (green, blue, purple). The density matrix of the initial state is an even-distributed diagonal matrix, which is the same as those in Fig.~3(a)(b)(c).
			} \label{FigE1}
		\end{figure}

	\end{document}